\documentclass[%
  reprint,
  superscriptaddress,
  preprintnumbers,
  nofootinbib,
  amsmath,amssymb,
  aps,
  prd,
]{revtex4-1}
\usepackage{psfrag,slashed,cancel,array,graphicx,hyperref}
\usepackage{orcidlink}
\usepackage{subfigure}
\usepackage[labelfont=bf,labelsep=quad,justification=justified]{caption}
\usepackage[utf8]{inputenc}
\usepackage{mathtools}
\usepackage{mathrsfs}
\usepackage{multirow}
\usepackage{braket}
\usepackage{slashed}
\usepackage{booktabs} %
\usepackage{slashed}
\hypersetup{pdftitle={Celestial energy-energy correlation in Yang-Mills theory and gravity},pdfcreator={},linkcolor=[rgb]{0.15,0.35,0.75},colorlinks=true,citecolor=[rgb]{0.675,0,0.2},urlcolor=[rgb]{0.15,0.35,0.65}}

\def\beq{
\begin{equation}}
  \def\eeq{
\end{equation}}
\def\bsp#1\esp{
  \begin{split}#1
\end{split}}

\newcommand{\tr}{\mathrm{Tr}}

\AtBeginDocument{
  \heavyrulewidth=.08em
  \lightrulewidth=.05em
  \cmidrulewidth=.03em
  \belowrulesep=.65ex
  \belowbottomsep=0pt
  \aboverulesep=.4ex
  \abovetopsep=0pt
  \cmidrulesep=\doublerulesep
  \cmidrulekern=.5em
  \defaultaddspace=.5em
}

\def\be{
\begin{equation}}
  \def\ee{
\end{equation}}

\newcommand{\cEEC}{\operatorname{cEEC}}
\newcommand{\Li}{\operatorname{Li}_2}

\begin{document}

\title{Celestial energy-energy correlation in Yang-Mills theory and gravity}

\author{HongYi Ruan\orcidlink{0009-0006-3233-922X}}
\email{ruanhongyi@stu.pku.edu.cn}
\affiliation{School of Physics, Peking University, Beijing 100871, China}
\author{YiZhe Zheng\orcidlink{0009-0000-8141-3234}}
\email{zhengyizhe@stu.pku.edu.cn}
\affiliation{School of Physics, Peking University, Beijing 100871, China}
\author{Hua Xing Zhu\orcidlink{0000-0002-7129-6748}}
\email{zhuhx@pku.edu.cn}
\affiliation{School of Physics, Peking University, Beijing 100871, China}
\affiliation{Center for High Energy Physics, Peking University, Beijing 100871, China}

\begin{abstract}
We introduce the celestial energy-energy correlator (cEEC), an infrared and collinear safe observable that makes the celestial conformal symmetry of four-dimensional scattering manifest. The cEEC is defined as a correlation function of average null energy operators measured on boost eigenstates and takes the form of a four-point function in a fictitious two-dimensional conformal field theory on the celestial sphere. An important feature of the cEEC is that it smoothly interpolates between different key regimes of perturbative gauge theory and gravity, such as the collinear limit, the Sudakov limit, and the Regge limit. We compute the cEEC to the first nontrivial order in $\mathcal{N}=4$ super Yang-Mills theory, pure Yang-Mills theory, Einstein gravity, and $\mathcal{N}=8$ supergravity. In $\mathcal{N}=8$ supergravity, the cEEC is uniquely determined by celestial symmetries and boundary data, demonstrating that bootstrap methods can yield closed-form results for this class of observables.
\end{abstract}

\maketitle

\section{Introduction}

In 1939, Dirac observed that the Lorentz group of 4D Minkowski space is isomorphic to the conformal group of the celestial sphere, suggesting a ``deep-lying connection'' between spacetime physics and complex analysis~\cite{dirac1939relation}. This vision was substantiated decades later by the discovery of the infinite-dimensional Bondi-Metzner-Sachs symmetry at null infinity~\cite{Bondi:1962px,Sachs:1962wk}, which revealed a rich asymptotic structure governing massless scattering. More recently, the celestial holography program~\cite{Pasterski:2016qvg,Strominger:2017zoo,Pasterski:2017kqt,Pasterski:2021raf,Pasterski:2017ylz,deBoer:2003vf,He:2015zea,Cheung:2016iub} has reorganized the S-matrix as correlation functions on the celestial sphere, recasting four-dimensional scattering amplitudes as two-dimensional conformal correlators.

While this program has revealed an intriguing web of connections between asymptotic symmetries, soft theorems, and memory effects~\cite{Strominger:2017zoo}, the construction of a complete holographic dictionary remains an active frontier. One challenge lies in the infrared divergences intrinsic to the massless $S$-matrix. To obtain well-defined celestial correlators, significant progress has been made by employing Faddeev-Kulish dressings~\cite{Kulish:1970ut,Kapec:2017tkm,Himwich:2020rro,Arkani-Hamed:2020gyp} to tame these singularities. Complementary to these amplitude-level approaches, it is also essential to bridge the gap between celestial symmetry and practical collider measurements. In collider physics, infrared finiteness is typically achieved not by dressing states, but by summing over final states to form inclusive cross sections~\cite{Kinoshita:1962ur,Lee:1964is}.

An important example of such infrared safe collider observables is the energy-energy correlator (EEC)~\cite{Basham:1978bw, Basham:1978zq}, a correlation function of the energy flow operator~\cite{Sterman:1975xv}. Originally introduced in the 1970s as a robust test of quantum chromodynamics, these correlators have recently undergone a renaissance, driven by their unique ability to bridge formal developments in quantum field theory~\cite{Hofman:2008ar,Belitsky:2013xxa,Kravchuk:2018htv} with precision collider physics~\cite{Dixon:2019uzg,Chen:2020vvp} and experiments~\cite{CMS:2024mlf,ALICE:2024dfl,CMS:2025ydi,ALICE:2025igw,Bossi:2024qeu,Electron-PositronAlliance:2025fhk,Liang-Gilman:2025gjl,Viinikainen:2025rix,Zhang:2025nlf,Electron-PositronAlliance:2025wzh,STAR:2025jut,Savoiu:2024uum,Connor:2024ehl,CMS-PAS-HIN-23-004} (see, e.g.,~\cite{Moult:2025nhu} for a review).

By itself, the energy flow operator, also known as the average null energy (ANE) operator~\cite{Sveshnikov:1995vi,Korchemsky:1999kt,Hofman:2008ar,Bauer:2008dt}, is a special case of light-ray operator~\cite{Kravchuk:2018htv} and has a long history; see, e.g.,~\cite{Hartman:2023qdn} and references therein. In particular, algebraic relations among the light-ray operators have been extensively studied~\cite{Cordova:2018ygx,Belin:2020lsr,Korchemsky:2021htm,Hu:2022txx,Hu:2023geb,Himwich:2025ekg,Sheta:2025oep}. However, a subtle issue remains: while the light-ray operator transforms simply, the standard EEC is defined as an expectation value in a momentum eigenstate and thus breaks the celestial symmetry down to $\mathrm{SO}(3)$ due to the choice of a specific frame. Recent work has shown that celestial symmetry emerges in the collinear limit of high-energy jets~\cite{Chen:2019bpb,Chen:2020adz, Chen:2021gdk,Chang:2022ryc,Chen:2022jhb,Chicherin:2024ifn}, but this restriction accesses only a fragment of the kinematic phase space, insensitive to other important regimes such as the Sudakov limit and the Regge limit.

In this paper, we propose the celestial energy-energy correlator (cEEC) by integrating the hadronic full-range EEC~\cite{Chen:2025rjc} over a measure that makes celestial symmetry manifest. The observable is infrared and collinear safe with respect to final-state radiation. Unlike previous constructions, the cEEC covers the full phase space, serving as a unified probe that interpolates between physically distinct regimes. It renders the operator product expansion limit, the double light-cone limit, and the Regge limit manifest within a single observable. We compute the cEEC to the first nontrivial order in Yang-Mills theory and gravity, revealing interesting zeros at the level of cross sections. We further show that in $\mathcal{N}=8$ supergravity, the result is uniquely fixed by the celestial symmetries and kinematic limits, suggesting that the cEEC is a promising target for the analytic bootstrap program.

\section{Definition of cEEC}
We start from the definition of full-range EEC~\cite{Chen:2025rjc},
\begin{equation}
  \label{eq:op def of ppEEC}
  \frac{d^2\sigma}{d\Omega_a\,d\Omega_b}=\frac{ \langle P_1 P_2 \left| \mathcal{E}(n_a) \mathcal{E}(n_b)\right| P_1 P_2\rangle }{2(P_1+P_2)^2\langle P_1 P_2 | P_1 P_2\rangle}\,,
\end{equation}
where $\mathcal{E}(n)$ is the ANE operator acting as an energy detector, and $|P_1 P_2\rangle$ is a two-particle incoming state with momenta $P_1$ and $P_2$. In hadron colliders, these are hadronic states with definite momentum.

This definition can be rewritten in a more useful representation by inserting a complete set of states labeled by collinear spin $J$ and additional quantum numbers $i$, with implicit summation:
\begin{equation}
  \label{eq:light-ray transition matrix}
  \frac{ \langle P_1 P_2 |\Psi_{J_1}^{(i_1)}\rangle\langle\Psi_{J_2}^{(i_2)}|P_1 P_2\rangle }{2(P_1+P_2)^2\langle P_1 P_2 | P_1 P_2\rangle}\langle\Psi_{J_1}^{(i_1)}| \mathcal{E}(n_a) \mathcal{E}(n_b)|\Psi_{J_2}^{(i_2)}\rangle\,.
\end{equation}
In the center of mass (c.m.) frame, $J$ is the eigenvalue of boosts along the beam axis. We refer to $|\Psi_{J_1}^{(i_1)}\rangle\langle\Psi_{J_2}^{(i_2)}|$ as a light-ray transition matrix, whose collinear spin is $J_1-J_2$.

As shown in Ref.~\cite{Chen:2025rjc}, choosing translation eigenstates as the initial state leads to a complicated structure because it generically probes transition matrices with different collinear spins, i.e., contributions with different $J_1-J_2$ mix. This motivates using boost eigenstates to project onto definite collinear spin.

We assume that $| P_1 P_2\rangle$ can be factorized as a tensor product of two-particle states $| P_1\rangle$ and $| P_2\rangle$~\footnote{Cases where this assumption might be false have been studied in \cite{Henn:2024qjq,Cieri:2024ytf,Duhr:2025lyg}.}. We can therefore treat the two beams independently. For massless particles, a boost eigenstate can be defined schematically as what we call a beam operator
\begin{equation}
  \label{eq: beam operator}
  \mathbb{P}^{(J)}(n)=\int_0^\infty \frac{dP}{P}\; P^{-J}\frac{|P\rangle\langle P|}{\langle P|P\rangle}.
\end{equation}
Here $P$ is a Lorentz scalar parameter and $|P\rangle$ is a one-particle state with momentum $P^\mu=P\,n^\mu$. Under a boost along $n^\mu$, $\Lambda_Y = e^{-i Y \vec n\cdot \vec{\mathbf{K}}}$, one finds
\begin{equation}
  \Lambda_Y \,\mathbb{P}^{(J)}(n)\,\Lambda_Y^{-1}
  =e^{J\,Y}\mathbb{P}^{(J)}(n)=\mathbb{P}^{(J)}(e^{Y}n).
\end{equation}

\begin{figure}[h]
  \centering
  \includegraphics[width=0.9\columnwidth]{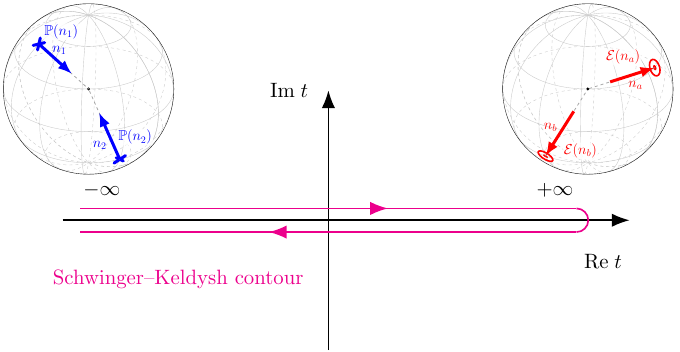}
  \caption{Schematic illustration of the operator definition of the cEEC on a Schwinger-Keldysh contour. The beam operators $\mathbb{P}_1^{J_1}$ and $\mathbb{P}_2^{J_2}$ prepare boost eigenstates as initial states, while the ANE operators $\mathcal{E}(n_a)$ and $\mathcal{E}(n_b)$ are inserted on the forward branch to measure energy flow in directions $n_a$ and $n_b$.}
  \label{fig: SK contour}
\end{figure}

Using these boost eigenstates as initial states, we define the cEEC as a correlator of ANE operators and beam operators on a Schwinger-Keldysh contour~\cite{Schwinger:1960qe, Haehl:2016pec}, see Fig.~\ref{fig: SK contour}:
\begin{equation}
  \label{eq: cEEC def}
  \frac{d^2\Sigma^{(J_1,J_2)}}{d\Omega_a d\Omega_b}
  =\tr \left[\mathcal{E}(n_a)\mathcal{E}(n_b)\mathbb{P}_1^{J_1}(n_1)\mathbb{P}_2^{J_2}(n_2)\right].
\end{equation}

Light-ray operators, including the ANE operator, are known to annihilate the vacuum~\cite{Kravchuk:2018htv,Kologlu:2019mfz}. In Eq.~\eqref{eq: cEEC def}, however, the contour-ordered trace with beam operators prepares a nonvacuum incoming state, and the correlator is therefore generically nonvanishing.

The cEEC is related to the full-range EEC in Eq.~\eqref{eq:op def of ppEEC} as
\begin{equation}
  \label{eq: cEEC EEC relation original}
  \frac{d^2\Sigma^{(J_1,J_2)}}{d\Omega_a d\Omega_b}=4 (n_1 \cdot n_2)\int_0^\infty dP_1 dP_2\;P_1^{-J_1}P_2^{-J_2}\frac{d^2\sigma}{d\Omega_a\,d\Omega_b}.
\end{equation}
Note that both the cEEC and the full-range EEC are Lorentz invariant. In fact, the full-range EEC takes the form~\cite{Chen:2025rjc}
\begin{equation}
  \label{eq: EEC form factor}
  \frac{d^2\sigma}{d\Omega_a\,d\Omega_b}=\frac{F(u,v,\hat w)}{4(n_a\cdot n_b)^3},
\end{equation}
where\footnote{Note that we are using a slightly different notation from the one used in \cite{Chen:2025rjc}.}
\begin{equation}
  \begin{aligned}
    \label{eq:u v w_hat using P}
    u=\frac{(P_2\cdot n_a)(P_1\cdot n_b)}{(P_1\cdot n_a)(P_2\cdot n_b)},    \quad v=\frac{(n_a\cdot n_b)(P_1\cdot P_2)}{(P_1\cdot n_a)(P_2\cdot n_b)},
  \end{aligned}
\end{equation}
are cross ratios that depend only on the directions $n_i$ since we parametrize the momenta as $P_i^\mu=P_i n_i^\mu$.
The dependence on $P_i$ enters only through
\begin{equation}
  \hat  w=\frac{P_1 \cdot n_a}{P_2\cdot n_a} \,,
\end{equation}
this motivates rewriting Eq.~\eqref{eq: cEEC EEC relation original} as
\begin{equation}
  \begin{split}
    \label{eq: cEEC form factor}
    \frac{d^3\Sigma^{(J_1,J_2)}}{d\Omega_a\, d\Omega_b\, dq^2}=\left(\frac{2n_1 \cdot n_2}{q^2}\right)^{\frac{J_1+J_2}{2}} \frac{w^{\frac{J_1-J_2}{2}}}{4(n_a\cdot n_b)^3}\;G^{(J_1,J_2)}(u,v),
  \end{split}
\end{equation}
where we have introduced $q^2 = (P_1 + P_2)^2$, $w = \frac{n_1 \cdot n_a}{n_2\cdot n_a}$, and
\begin{equation}
  G^{(J_1,J_2)}(u,v) =  \int_{0}^{\infty} \frac{d\hat w}{\hat w}\; \hat w^{-\frac{J_1 - J_2}{2}} \, F\left(u, v, \hat w\right) \,.
  \label{eq: G}
\end{equation}
In the c.m. frame, $\hat  w = e^{-2\eta_a}$ with $\eta_a$ the rapidity of detector $a$, and Eq.~\eqref{eq: G} can be written as
\begin{equation}
  G^{(J_1,J_2)}(u,v) = 2 \int_{-\infty}^{+\infty} d\eta_a \; e^{(J_1 - J_2)\eta_a} \, F\left(u, v, e^{-2\eta_a}\right) \,.
  \label{eq: integral}
\end{equation}
Here we are considering an ideal detector with infinite rapidity coverage.

Remarkably, with boost eigenstates as the initial states, the cEEC depends on $w$ only through a power law prefactor. By contrast, the full-range EEC has a nontrivial dependence on the analogous variable $\hat w$. This parallels the simplification of moments of parton distribution functions (PDFs)~\cite{Catani:1994sq,Lipatov:1996ts,Li:2025knf}. We may view the cEEC as a moment transform of the full-range EEC, which isolates contributions of light-ray transition matrices with different collinear spin $J_1-J_2$. As a consequence, it inherits several properties of EEC, including infrared and collinear safety~\cite{Basham:1978bw,Basham:1978zq} and positivity~\cite{Faulkner:2016mzt,Hartman:2016lgu}.

More importantly, up to the trivial dependence on the c.m. energy, this observable depends only on the four future-pointing null vectors that label the operators in Eq.~\eqref{eq: cEEC def}. These vectors lie on the future light cone, which can be viewed as the embedding space of the 2D celestial sphere~\cite{Dirac:1936fq,deAlfaro:1976vlx}. Accordingly, the cEEC behaves like a fictitious four-point correlator in a Euclidean 2D conformal field theory with global conformal symmetry, with $\mathcal{E}(n)$ and $\mathbb{P}^J(n)$ playing the role of primary operators of celestial dimensions $3$ and $-J$.

\section{Explicit expressions and general properties}

We now explain how to evaluate the formally defined cEEC in perturbation theory for quark/gluon and graviton scattering. We begin with the full-range EEC in Eq.~\eqref{eq:op def of ppEEC}. Inserting a complete set of states $|X\rangle\langle X|$ between the two ANE operators yields a Lorentz invariant representation,
\begin{equation}
  \begin{aligned}
    \label{eq:Lorentz invariant form for ppEEC}
    \frac{d^2\sigma}{d\Omega_a d\Omega_b} =\frac{1}{8\pi ^2q^2(n_a\cdot q)^2(n_b\cdot q)^2}\sum_{n=2}^{\infty}\sum_k \int\frac{d\Pi_n}{\text{sf}_{n,k}}\\
    \overline{\sum \left| \mathcal{A}_{2+n,k} \right|^2}
    \sum\limits_{i,j=3}^{n+2}
    (p_i\cdot q)^2 (p_j\cdot q)^2
    \delta(p_i\cdot n_a)\delta(p_j\cdot n_b),
  \end{aligned}
\end{equation}
where $q^\mu=P_1^\mu+P_2^\mu$ is the total momentum. Here $n$ is the number of final-state particles and $k$ labels the contributing diagrams at fixed $n$. The factor $\text{sf}_{n,k}$ is the corresponding symmetry factor, and $\overline{\sum \left| \mathcal{A}_{2+n,k} \right|^2}$ denotes the averaged squared amplitude of $2 \to n$ scattering. Here we are summing over the final-state species, as is standard in inclusive observables. The measure $d\Pi_n$ is the standard Lorentz invariant $n$ body phase space measure.

To extract the form factor $F(u,v,\hat w)$ in Eq.~\eqref{eq: EEC form factor}, it suffices to evaluate Eq.~\eqref{eq:Lorentz invariant form for ppEEC} in the c.m. frame, where $q^\mu$ is purely timelike and $n_a^\mu=(1,\;\vec n_a)$, $n_b^\mu=(1,\;\vec n_b)$.

In the Mathematica notebook~\texttt{cEEC\_Results.nb}~\cite{Zenodo:cEEC}, we provide the leading-order full-range EEC in the c.m. frame for several theories, including $\mathcal{N}=8$ supergravity (SUGRA), pure Einstein gravity, $\mathcal{N}=4$ super Yang-Mills (SYM) theory, and pure Yang-Mills theory.\nocite{Maitre:2005uu,Maitre:2007kp} We refer to \cite{Gonzo:2020xza,Herrmann:2024yai,Chicherin:2025keq} for related studies of energy correlators in gravitational theories.
The full-range EEC first receives a nontrivial contribution from five-point tree amplitudes, which are available in the literature, for example, in Ref.~\cite{Elvang:2013cua}.

\begin{figure}[h]
  \centering
  \includegraphics[width=0.7\columnwidth]{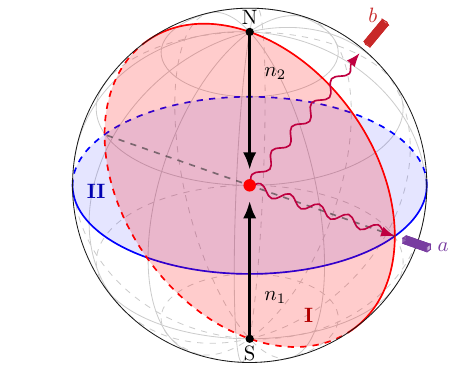}
  \caption{Standard configuration of Eq.~\eqref{eq: std configuration}, with detector $b$ at an arbitrary point on the sphere.}
  \label{fig: standard_configuration}
\end{figure}

Since our observable is fully Lorentz invariant, it is convenient to work in a standard configuration as illustrated in Fig.~\ref{fig: standard_configuration}.
Parameterizing in stereographic coordinates, this corresponds to
\begin{equation}
  \label{eq: std configuration}
  z_1=0,\quad z_2=\infty,\quad z_a=1,
\end{equation}
leaving $z_b=z$ and its complex conjugate $\bar z$ as the only independent kinematic variables. In this frame, the cross ratios reduce to
\begin{equation}
  u=z\bar z,\qquad v=(1-z)(1-\bar z) \,,
\end{equation}
where the Lorentz symmetry acts as the global conformal group on the complex plane.

\begin{figure}[h]
  \centering
  \includegraphics[width=0.9\columnwidth]{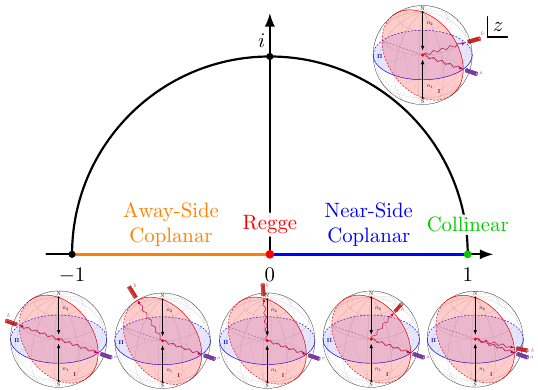}
  \caption{All information of the cEEC is contained in the upper half unit disk, with a direct physical interpretation.}
  \label{fig: cEEC_half_unit_disc}
\end{figure}

For identical incoming beams, the standard configuration of Fig.~\ref{fig: standard_configuration} admits two parity symmetries, given by reflections about planes~I and~II, which act as $z\leftrightarrow \bar z$ and $z\leftrightarrow 1/\bar z$, respectively. The first implies the reality of the cEEC, and together they restrict the independent kinematics to the upper half unit disk shown in Fig.~\ref{fig: cEEC_half_unit_disc}.

In this configuration, Eq.~\eqref{eq: cEEC form factor} reduces to
\begin{equation}
  \begin{split}
    \label{eq: cEEC_std}
    \cEEC^{(J_1, J_2)}(z, \bar z)=& \left(\frac{4}{q^2}\right)^{\frac{J_1+J_2}{2}}\frac{ (1+z\bar z)^3 G^{(J_1, J_2)}(u,v)}{4 (1-z)^3(1-\bar z)^3},
  \end{split}
\end{equation}

The rapidity integral defining $G^{(J_1,J_2)}(u,v)$ may develop endpoint divergences from $\eta_a\to\pm\infty$, depending on the collinear behavior. For generic values of $u,v$, these endpoints correspond to both detectors becoming nearly aligned with the beam, with the opening angle $\chi$ scaling as $e^{-|\eta_a|}$. In gravity, there is no collinear singularity, and the form factor $F$ is exponentially suppressed at large rapidities, $F\sim e^{-2|\eta_a|}$ as $|\eta_a|\to\infty$. By contrast, in Yang-Mills theory the collinear divergence $\chi^{-2}$ offsets this suppression, giving $F\sim \mathcal{O}(1)$ at both end points.

Therefore, for $J_1=J_2$ (collinear spin zero transition matrix), the integral over $\eta_a$ is finite in gravity but linearly divergent in Yang-Mills theory.  In this work we focus on $J_1=J_2$, while generalizations to other collinear spins remain possible.
Finally, since the prefactor $\left(4/q^2\right)^{\frac{J_1+J_2}{2}}$ in Eq.~\eqref{eq: cEEC_std} will not play an essential role, we will set $J_1=J_2=0$ in what follows. 

We now present explicit cEEC results and discuss how different limits can be accessed, with emphasis on $\mathcal{N}=8$ SUGRA. Additional results are collected in the Appendix and in the Mathematica notebook~\texttt{cEEC\_Results.nb}~\cite{Zenodo:cEEC}.

At the first nontrivial order, the cEEC in $\mathcal{N}=8$ SUGRA takes a remarkably simple form,
\begin{equation}
  \begin{aligned}
    \label{eq: ceec_sugra}
    \cEEC^{(0,0)}_{\text{SUGRA}}(z,\bar z)= c_0\frac{\left(1+z \bar{z}\right)^3}{z\bar z(z-\bar z)}\left(
      \frac{\bar z}{1-\bar{z}}\ln ^2(z)\right. \\
      \left.
      +\frac{z-\bar z}{2  }\frac{\ln (z)}{1-z} \frac{\ln (\bar{z})}{1-\bar{z}}
      -2 \Li(1-z)
    \right)+(z\leftrightarrow \bar z)\,,
  \end{aligned}
\end{equation}
where
\begin{equation}
  c_0=\frac{(\kappa q)^6 }{(4 \pi)^5},\qquad \kappa^2=8\pi G_N,
\end{equation}
with $q$ the c.m. energy and $G_N$ the Newton constant. Equation~\eqref{eq: ceec_sugra} should be understood on the principal branch, with a branch cut along the negative real axis. It would be interesting to explore the physical meaning of its analytic continuation to other branches. In what follows, we drop the overall factor $c_0$.

\begin{figure}[h]
  \centering
  \includegraphics[width=0.9\columnwidth]{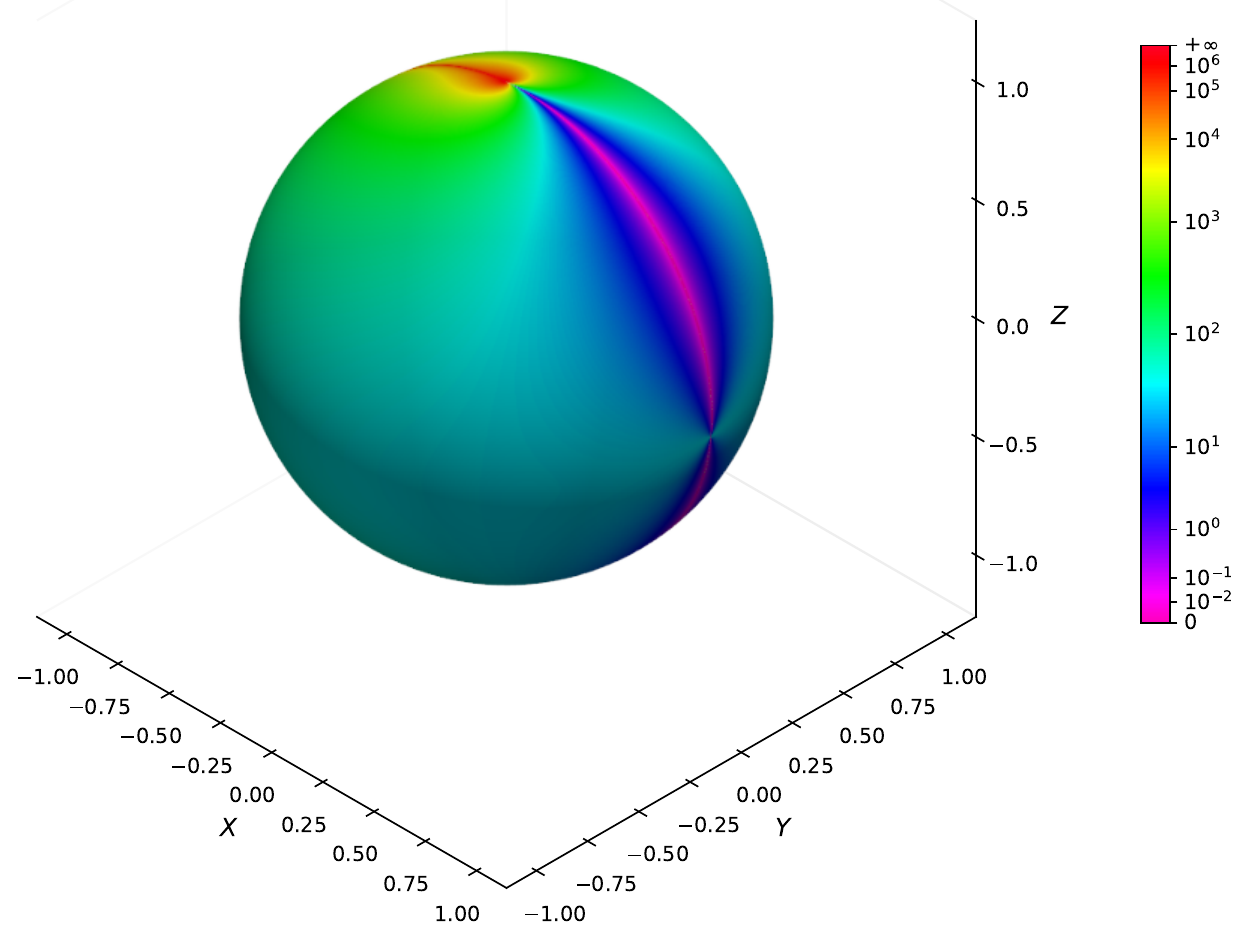}
  \caption{$\cEEC^{(0,0)}_{\text{SUGRA}}(z,\bar z)$ on the celestial sphere. Detector $a$ is fixed at $(1,0,0)$, while the color map indicates the value of the cEEC as detector $b$ moves around the sphere.}
  \label{fig: cEEC_sugra}
\end{figure}
Using stereographic coordinates, we visualize $\cEEC^{(0,0)}_{\text{SUGRA}}$ on the celestial sphere in Fig.~\ref{fig: cEEC_sugra}. The kinematic limits of interest lie on the real axis of the $z$ plane, corresponding to the great circle passing through the three fixed points on the sphere.

To study these limits systematically, we introduce polar coordinates on the $z$ plane,
\begin{equation}
  \label{eq:polar coordinates}
  z=r\, e^{i\psi},\quad \bar z=r\,e^{-i\psi},
\end{equation}
where $\psi$ is the azimuthal angle of detector $b$ with respect to the beam axis in the standard configuration.

In the Regge limit $r\to 0$, the most singular terms in $\mathcal{N}=8$ SUGRA and pure Einstein gravity coincide,
\begin{equation}
  \label{eq: regge limit}
  \cEEC^{(0,0)}_{\text{(SU)GRA}}\xrightarrow{r\to0} 2\,
  (\psi  \cot (\psi )-1)\frac{ \ln r }{r^2}.
\end{equation}
The nontrivial $\psi$ dependence implies qualitatively different behaviors on the near side ($\psi\to 0$) and on the away side ($\psi\to \pi$), which will reappear in the coplanar limits.

A key observation is that, at this perturbative order in gravity, the cEEC vanishes identically on the whole positive real axis, which corresponds to the near-side coplanar line, except for the collinear point $z=\bar z=1$. This observable-level null line is reminiscent of the hidden zeros of scattering amplitudes discussed in Refs.~\cite{Cachazo:2021wsz,Arkani-Hamed:2023swr,Cao:2024gln,DAdda:1971wcy,Bartsch:2024amu,Li:2024qfp,Guevara:2024nxd,Chang:2025cqe,Rodina:2024yfc,Cao:2024qpp,Zhou:2024ddy,Zhang:2024efe,Li:2024bwq,Huang:2025blb,Zhou:2025tvq,Zhang:2025zjx,De:2025bmf,Feng:2025ofq,Jones:2025rbv,Li:2025suo,Backus:2025hpn,CarrilloGonzalez:2026lnu}. With this vanishing established, the positivity of the cEEC then implies that the first nonvanishing perturbative correction is constrained to be non-negative along this line, and may therefore provide a useful starting point for deriving positivity bounds in quantum gravity.

\begin{figure}[h]
  \centering
  \includegraphics[width=0.9\columnwidth]{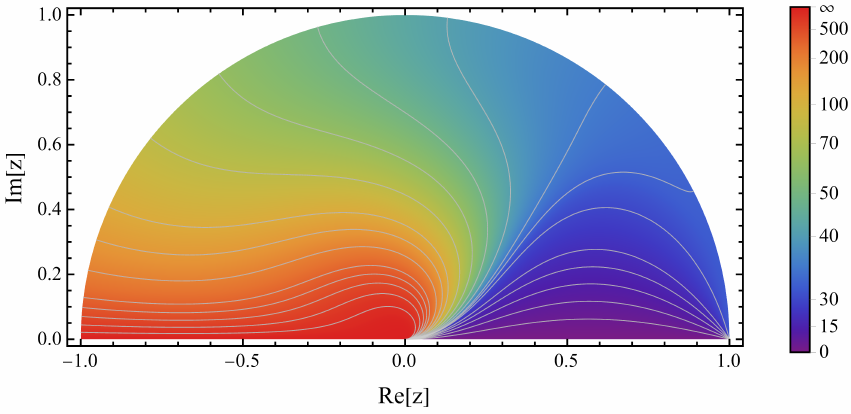}
  \caption{$\cEEC^{(0,0)}_{\text{SUGRA}}(z,\bar z)$ on the upper half unit disk in the complex plane. One can clearly identify the Regge and away-side coplanar singularities, as well as the zero on positive real axis.}
  \label{fig: cEEC_SUGRA_cp}
\end{figure}

Moving slightly off this null line, the cEEC exhibits a pronounced enhancement near the collinear point $z=\bar z=1$.
To probe this collinear behavior, we focus on the vicinity of $z=1$ by reparametrizing
\begin{equation}
  z=1+\tilde r\,e^{i\tilde \psi},\quad \bar z=1+\tilde r\,e^{-i\tilde \psi},
\end{equation}
followed by the limit $\tilde r\to 0$. In this parametrization, the leading collinear behavior reads
\begin{equation}
  \label{eq: collinear limit}
  \cEEC^{(0,0)}_{\text{SUGRA}}\xrightarrow{\tilde r\to0}32\sin^2\tilde \psi,\quad
  \cEEC^{(0,0)}_{\text{GRA}}\xrightarrow{\tilde r\to0}\frac{968}{81}\sin^2\tilde \psi.
\end{equation}
Here $\tilde\psi$ is the azimuthal angle around $n_a$. The characteristic $\sin^2\tilde \psi$ dependence reflects the spin two nature of graviton exchange, as also noted in Ref.~\cite{Herrmann:2024yai}.

In the away-side coplanar limit $\psi\to \pi$, both theories exhibit a universal enhancement proportional to $1/(\pi-\psi)$. We note that the strict back-to-back limit $r\to1$ is not more singular than a generic point on this coplanar line.

To make contact with Fig.~\ref{fig: cEEC_half_unit_disc} and to exhibit the kinematic limits (including the near-side null line) more transparently, we also plot $\cEEC^{(0,0)}_{\text{SUGRA}}(z,\bar z)$ on the upper half unit disk in the complex plane in Fig.~\ref{fig: cEEC_SUGRA_cp}, where these features are clearly visible.

For $\mathcal{N}=4$ SYM theory and pure Yang-Mills theory, the integral in Eq.~\eqref{eq: integral} is divergent. To display this explicitly, we introduce a simple symmetric cutoff,
\begin{equation}
  \label{eq: YM regulator eta}
  G^{(J_1,J_2)}(u,v)=\lim\limits_{Y_0\to \infty} 2\int_{-Y_0/2}^{Y_0/2} d\eta_a\, e^{(J_1-J_2)\eta_a} F(u,v,e^{-2\eta_a}) \,,
\end{equation}

For $\mathcal{N}=4$ SYM theory, the divergent part reads
\begin{equation}
  \begin{split}
    \cEEC^{(0,0)}_{\text{SYM, div}}=\frac{8g^6 N_c^3  }{(4 \pi) ^5
    \left(N_c^2-1\right)}\frac{(1+z \bar z)^3 }{  z \bar z (1-z) (1-\bar z)  (z-\bar z)} \\
    (2 z \bar z-z-\bar z+2)(\ln (z)-\ln (\bar z))  Y_0.
  \end{split}
\end{equation}
The divergence indicates that the initial state defined by the beam operator in Eq.~\eqref{eq: beam operator} requires renormalization, analogous to the renormalization of detector operators~\cite{Caron-Huot:2022eqs, Chang:2025zib, Chen:2023zzh, Chen:2019bpb, Chen:2020vvp, Chen:2020uvt, Chen:2020adz, Chen:2021gdk}. We defer a systematic treatment to future work and collect the remaining expressions in the Appendix and in the Mathematica notebook~\texttt{cEEC\_Results.nb}~\cite{Zenodo:cEEC}.

\section{Analytic bootstrap}
There have been recent attempts to apply bootstrap methods to EEC type observables~\cite{Gong:2025jqi}. The challenge is that ordinary EEC observables depend on complicated rational functions, in addition to the usual transcendental functions. By contrast, the cEEC in Eq.~\eqref{eq: ceec_sugra} exhibits a strikingly simple structure. In this section, we show that the result in Eq.~\eqref{eq: ceec_sugra} can be efficiently reconstructed using known limit data and symmetry constraints.

The logarithms in the kinematic limits suggest the alphabet $\{z,\bar z,1-z,1-\bar z\}$. The same limits indicate a denominator at least as singular as $z\bar z(1-z)(1-\bar z)(z-\bar z)$, with bounded polynomial dependence in the numerator. Imposing the $z\leftrightarrow\bar z$ symmetry, we write a schematic ansatz as
\begin{equation}
  \begin{aligned}
    \frac{(1+z\bar z)^3}{z\bar z(1-z)(1-\bar z)}\left[\sum_{\text{even}}\left(\sum_{n,m}a_{m,n}(z+\bar z)^m(z\bar z)^n\right)\text{FB}_{\text{even}}\right.\\
    \left.+\sum_{\text{odd}}\left(\sum_{n,m}b_{m,n}(z+\bar z)^m(z\bar z)^n\right)\frac{\text{FB}_{\text{odd}}}{z-\bar z}\right]
  \end{aligned}
\end{equation}
where $(z-\bar z)$ appears only in the odd sector.

The strongest constraint is inversion symmetry. Using the inversion action on the transcendental basis yields linear constraints on $a_{m,n}$ and $b_{m,n}$, leaving $13$ parameters after scanning $m\le 10$ and $n\le 5$. Matching a minimal set of limit data then fixes the result in Eq.~\eqref{eq: ceec_sugra}. One choice uses the collinear behavior in Eq.~\eqref{eq: collinear limit} and the leading away-side coplanar singularity given in the Appendix. A second, more accessible choice uses the absence of a collinear divergence and the leading Regge behavior in Eq.~\eqref{eq: regge limit}. We implement this bootstrap in the Mathematica notebook~\texttt{cEEC\_Bootstrap.nb}~\cite{Zenodo:cEEC} using \texttt{POLYLOGTOOLS}~\cite{Duhr:2019tlz}.

\section{Outlook}

In this work, we have introduced the cEEC, a correlation function of ANE operators in boost eigenstates. The cEEC inherits many properties of the conventional EEC, while exhibiting rich structure. We have presented explicit results in gravitational and gauge theories and demonstrated the possibility to reconstruct the cEEC using analytic bootstrap methods. There are a number of future avenues to explore. First, it would be interesting to compute higher-order corrections in perturbation theory for both gauge and gravitational theories. Second, it would be interesting to understand the singular structures in various limits~\cite{Chicherin:2025keq}. Third, the cEEC may provide a natural observable for understanding the implications of asymptotic symmetries and infrared structure in gauge and gravitational theories, see, e.g.,~\cite{Gonzalez:2025ene,Moult:2025njc}. We leave them for future work.

\section{Acknowledgments}
We thank Qu Cao, Zhongjie Huang, Bo Wang, and Xiaoyuan Zhang for useful discussions. We are especially grateful to Hao Chen for insightful discussions at an early stage of this work. This work is supported by the National Natural Science Foundation of China under Grant No.~12425505. H.X.Z. would also like to thank the Southern Center for Nuclear-Science Theory (SCNT), Institute of Modern Physics, Chinese Academy of Sciences for hospitality while this work was carried out.

\section*{Data Availability}
The Wolfram Mathematica notebooks supporting the findings of this study, including analytic expressions, intermediate results, and the analytic bootstrap reconstruction, are publicly available~\cite{Zenodo:cEEC}. No external datasets were used in this study.

\bibliographystyle{apsrev4-1}
\bibliography{refs}{}

\onecolumngrid
\appendix
\section{Celestial EEC for other theories}

We collect here additional plots, limiting behaviors, and expressions that are not included in the main text.

\begin{figure}[t]
  \centering
  \includegraphics[width=0.8\columnwidth]{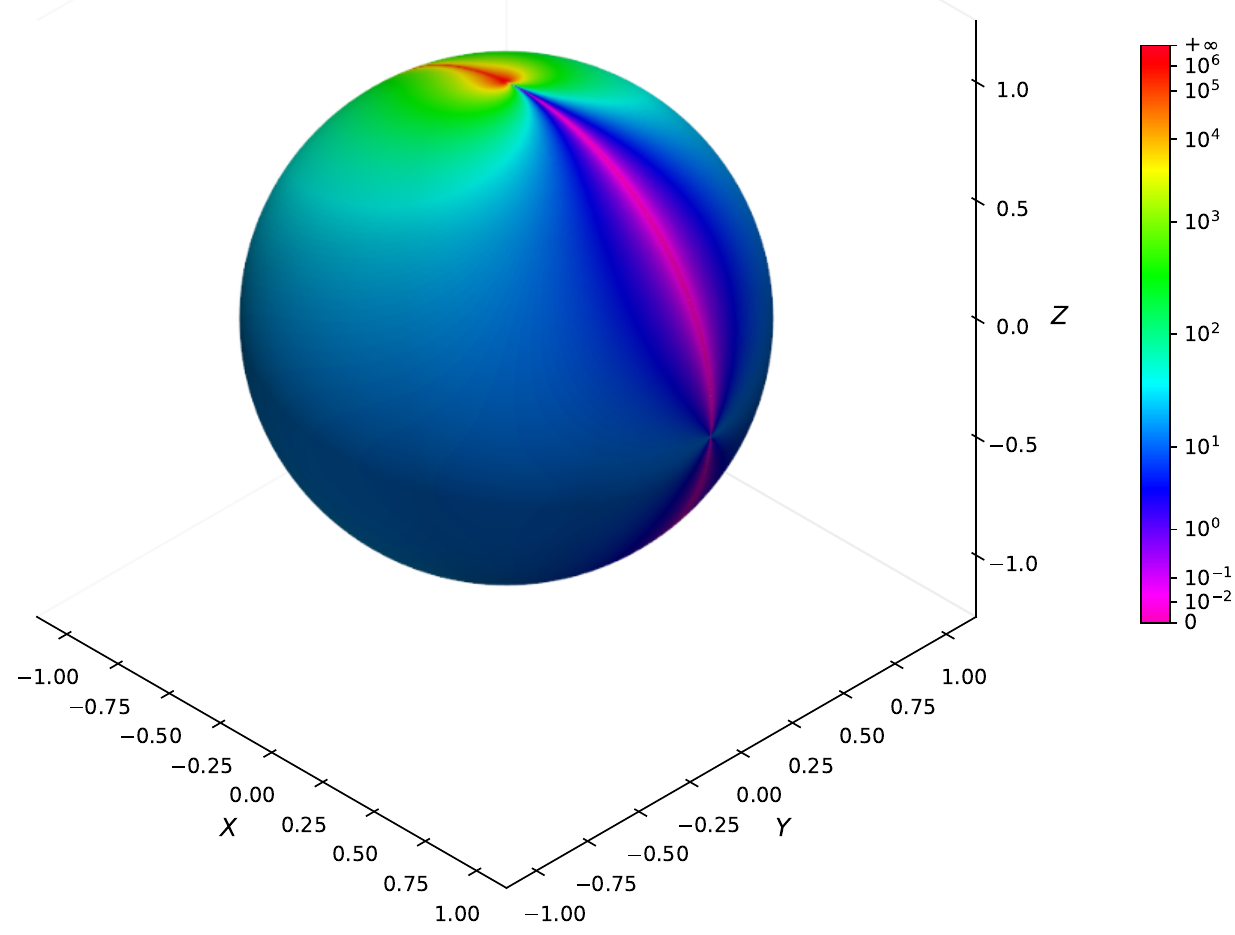}
  \caption{$\cEEC^{(0,0)}_{\text{GRA}}(z,\bar z)$ on the celestial sphere, shown as a color map.}
  \label{fig: cEEC_gra}
\end{figure}

Using the Einstein gravity result $\cEEC^{(0,0)}_{\text{GRA}}$ provided in the Mathematica notebook~\texttt{cEEC\_Results.nb}~\cite{Zenodo:cEEC}, we map the expression from the complex plane back to the celestial sphere and obtain Fig.~\ref{fig: cEEC_gra}. The same kinematic limits discussed in the main text are visible, including the Regge, collinear, near-side coplanar, and away-side coplanar limits.

As discussed in the main text, the leading Regge behavior of $\cEEC^{(0,0)}_{\text{GRA}}$ is identical to that of $\cEEC^{(0,0)}_{\text{SUGRA}}$ and is given in Eq.~\eqref{eq: regge limit}. Expanding Eq.~\eqref{eq: regge limit} in the angular variable $\psi$ makes the near- and away-side dependence more explicit:
\begin{equation}
  \cEEC^{(0,0)}_{\text{(SU)GRA}}
  \xrightarrow[\psi\to0]{r\to0}
  -\frac{2}{3}\,\psi^2\,\frac{\ln r}{r^2},
  \qquad
  \cEEC^{(0,0)}_{\text{(SU)GRA}}
  \xrightarrow[\psi\to\pi]{r\to0}
  -\frac{2\pi}{\pi-\psi}\,\frac{\ln r}{r^2}.
\end{equation}

We further highlight the near-side coplanar behavior in pure Einstein gravity.
At this perturbative order, $\cEEC^{(0,0)}_{\text{GRA}}$ vanishes identically along the entire positive real axis, except at the collinear point $z=\bar z=1$, giving a sharp, observable-level null line.

\begin{figure}[t]
  \centering
  \includegraphics[width=0.8\columnwidth]{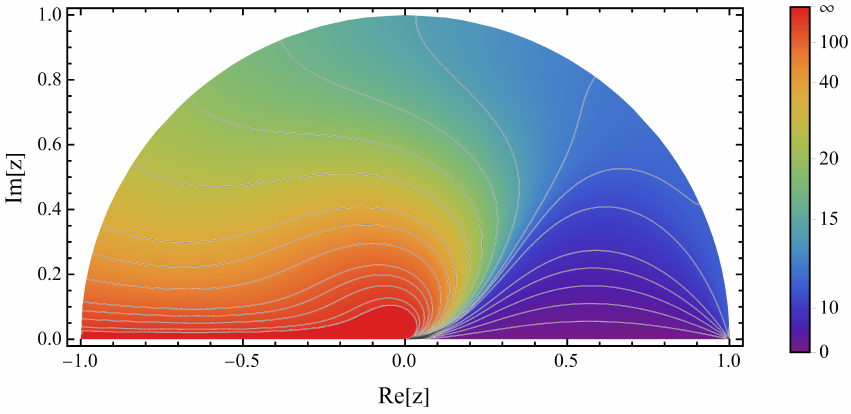}
  \caption{$\cEEC^{(0,0)}_{\text{GRA}}(z,\bar z)$ on the upper half unit disk in the complex plane.}
  \label{fig: cEEC_gra_cp}
\end{figure}

In the away-side coplanar limit $\psi\to\pi$, both $\mathcal{N}=8$ SUGRA and pure Einstein gravity show an enhancement proportional to $1/(\pi-\psi)$. Moreover, the back-to-back point $r\to1$ is not parametrically more singular than a generic point on this coplanar line. For completeness, the leading away-side coplanar behavior is
\begin{equation}
  \begin{aligned}
    \cEEC^{(0,0)}_{\text{SUGRA}} \xrightarrow{\psi\to\pi}
    \frac{2 \pi  \left(r^2+1\right)^3 ((r+1) \ln (r+1)-r \ln (r))}{r^3 (r+1) (\pi -\psi )},
  \end{aligned}
\end{equation}
and
\begin{equation}
  \begin{aligned}
    \cEEC^{(0,0)}_{\text{GRA}} \xrightarrow{\psi\to\pi}
    \frac{\pi r \left(r^2+1\right)^3 ((r+1) \ln (r+1)-r \ln (r))}{(r+1)^9 (\pi -\psi )}
    \left(2r^4+8 r^3+28 r^2+56 r+35 \right)+\left(r\leftrightarrow r^{-1}\right).
  \end{aligned}
\end{equation}
For completeness, we also visualize $\cEEC^{(0,0)}_{\text{GRA}}(z,\bar z)$ on the complex plane in Fig.~\ref{fig: cEEC_gra_cp}. While it carries essentially the same information as the sphere representation in Fig.~\ref{fig: cEEC_gra}, the complex-plane plot makes the limiting behaviors more transparent.

Finally, we record representative expressions in Yang-Mills theories beyond those shown in the main text. The regular term of the cEEC in $\mathcal{N}=4$ SYM theory is
\begin{equation}
  \begin{aligned}
    \cEEC^{(0,0)}_{\text{SYM,reg}}(z,\bar z)=\frac{g^6 N_c^3 (z \bar z+1)^3 }{512 \pi ^5 \left(N_c^2-1\right) z \bar z (1-z) (1-\bar z) (\bar z-z)}\bigg(
      2 \Li(1-z) (5z \bar z-2z-2 \bar z+5)\\
      -(5 z \bar z-3 z-3 \bar z+6)\ln ^2(z)
      +4(2z\bar z-z-\bar z+2)\left(\Li\left(\frac{z}{\bar z}\right)+\ln (z) \ln (z-\bar z) +\ln (z) \ln (\bar z-z) \right)
    -(z\leftrightarrow \bar z) \bigg),
  \end{aligned}
\end{equation}
and the divergent term in pure Yang-Mills theory is
\begin{equation}
  \begin{aligned}
    \cEEC^{(0,0)}_{\text{YM,div}}(z,\bar z)=\frac{g^6 N_c^3 (z \bar z+1)^3}{256 \pi^5 \left(N_c^2-1\right)} \frac{2 z \bar z-z-\bar z+2}{z \bar z(z-1)(\bar z-1)}
    \left(\frac{2\left(z^2-z+1\right)^2  \ln (z)}{(z-1)^4(z-\bar z)}\right.\\
    \left.+\frac{7 z^2 \bar z^2-19 z^2\bar z+11 z^2+13 z \bar z-19 z+7}{3 (z-1)^3 (\bar z-1)^3 }+(z\leftrightarrow \bar z)\right)Y_0.
  \end{aligned}
\end{equation}
Additional expressions are provided in the Mathematica notebook~\texttt{cEEC\_Results.nb}~\cite{Zenodo:cEEC}.

\end{document}